\documentclass{ws-ijmpd}

\begin{document}

\markboth{P. K. Chattopadhyay, R. Deb, B. C. Paul}
{Relativistic Solution for a Class ...}

%
\catchline{}{}{}{}{}
%

\title{Relativistic Solution for a Class of Static Compact Charged  Star in Pseudo Spheroidal Space-Time}

\author{Pradip Kumar Chattopadhyay\footnote{$pkc_{-}76$@rediffmail.com}}

\address{Department of Physics, Alipurduar College,\\
P.O.: Alipurduar Court, Dist: Jalpaiguri, Pin: 736122,\\West Bengal, India\\}

\author{Rumi Deb}

\address{IRC, Physics Department, North Bengal University, Dist: Darjeeling,\\Pin: 734013,
West Bengal, India }

\author{Bikash Chandra Paul\footnote{bcpaul@iucaa.ernet.in}}

\address{Department of Physics, North Bengal University, Dist: Darjeeling,\\Pin: 734013,
West Bengal, India }

\maketitle

\begin{history}
\received{}
\revised{}
\comby{}
\end{history}

\begin{abstract}
Considering Vaidya-Tikekar metric, we obtain  a class of solutions of the  Einstein-Maxwell equations for a charged static fluid sphere. The physical 3-space (t=constant) here is described by   pseudo-spheroidal geometry. The relativistic solution for the theory is used to obtain models for charged compact objects, thereafter a qualitative analysis of the physical aspects of  compact objects are studied. The dependence of some of the properties of a superdense star on the parameters of the three geometry is explored.   We note that the  spheroidicity parameter $a$, plays an important role for determining the properties of a compact object. A non-linear equation of state is required to describe  a charged compact object with pseudo-spheroidal geometry which we have shown for known masses of compact objects. We also note that the size of a  static compact charged star is more than that of a static compact star without charge.
\end{abstract}


\section{Introduction}

The analysis of very compact astrophysical objects has been a key issue in relativistic astrophysics for the last few  years. The estimated mass and radius of several compact objects such as X-ray pulsar Her X-1, X-ray burster 4U 1820-30, millisecond pulsar SAX J 1808.4-3658, X-ray sources 4U 1728-34, PSR 0943+10 and RX J185635-3754, are not compatible with the standard neutron star models. There are several astrophysical objects as well as cosmological phases where one needs to consider equation of state of matter    involving matter densities of the order of  $10^{15}\; gm/cc$ or higher, exceeding the nuclear matter density. The conventional approach of obtaining models of relativistic stars in equilibrium requires a definite information about the equation of state of its matter content. The equation of state for a superdense star is yet to be determined. An alternative approach generally followed\cite{Vaidya,RT1,Mukherjee}\cdash\cite{SK}  in relativity is to prescribe a suitable ansatz geometry for the interior three space in order to extract physical properties of  such stars.
In relativistic astrophysics, exact solutions of Einstein-Maxwell equations for the interior matter content in the form of charged perfect fluid spheres become important because of its various physical applications\cite{RS02,RS04}. The solution of the interior space-time of a charged sphere is matched with the exterior Reissner-Nordstrom metric at the boundary. In the presence of charge, the gravitational collapse may be counter balanced by the electrostatic repulsive Columbian force along with the pressure gradient. As a result gravitational collapse of a spherically symmetric matter distribution with charge might avoid collapse to a point singularity. Recently a number of papers in stellar models came up considering electromagnetic field\cite{RT,PKC,RS02}. The matter density of many compact objects are normally above the nuclear density. It has maximum mass and radius both less than that of neutron stars, with higher compactification factor (ratio of mass to radius). Compact objects are classified on the basis of compactification factor, for a normal stars $(10^{-5})$, white dwarfs $(10^{-3})$, neutron stars (0.1 to 0.2), strange stars (0.2 to $<$ 0.5), black holes $(0.5)$ etc. The physics near the core region of a super-dense compact object is not well understood yet. The equation of state of compact objects such as neutron stars, strange stars is also not known at least near the core region. Therefore a compact object can be analysed by considering a simple spatial geometry characterised by spheroidicity parameter $(a)$ and curvature parameter $(R)$. In this context, Vaidya and Tikekar\cite{Vaidya} and Tikekar\cite{RT1} prescribed a simple form for the space-like hyper-surface (t=constant) containing two such parameters namely, spheroidicity parameter $(a)$ and curvature parameter $(R)$. Vaidya-Tikekar approach reduces the complexity of the field equations which produces solution of relativistic stars with ultra-high densities and pressures that is consistent with the observations\cite{RCE}. The Vaidya-Tikekar approach is also useful to obtain stellar solution for a compact star with Einstein-Maxwell field equations. The solution is characterised by two parameters $a$ and $R$. The physical properties of static charged star is studied in different context \cite{PLK,RT,Patel} considering Vaidya-Tikekar space time metric\cite{Vaidya}.  The physical plausibility of Vaidya-Tikekar approach is discussed in the literature\cite{Vaidya,RT1,Knutsen}. Maharaj and Leach\cite{ml} analyzed a relativistic model for superdense stars proposed by Tikekar\cite{RT1} extensively and obtained new classes of solutions for different choice of the spheroidal parameter.
Tikekar and Thomas\cite{Thomas},\cite{VT}  analysed compact stars with 3-pseudo-spheroidal geometry for the 3-space of the interior space-time and obtained a class of relativistic solutions suitable for modelling. In this paper, we obtain a class of relativistic stellar models using Tikekar-Thomas ansatz which prescribes 3-pseudo spheroidal geometry for the 3-space for the interior space-time in the presence of charge. A number of papers appeared in the literature\cite{Thomas,pmehta,VT,Tikekar}
considering  pseudo-spheroidal geometry to construct compact star models including anisotropic distribution of fluid.  The suitability of the solution for describing the model of a superdense star in pseudo-spheroidal geometry may be important to explore. Usually knowing equation of state (henceforth, EOS) we determine the geometry from the Einstein's field equation but as the EOS of a neutron star is unknown so we adopt a different technique\cite{Mukherjee} to determine  the EOS assuming a pseudo-spheroidal space-time inside a compact star. Using the metric, we establish a relation for pressure and density in the Einstein gravity and study physical properties of a suprdense star.\\
The outline of the paper is as follows. In sec. 2, we set up the relevant field equation and obtain relativistic solution for different configuration with total charge. In sec. 3, we determine the constraints under which the solutions are physically realistic. Thereafter the physical properties of compact object for known mass and radius are explored. The parameters $a$ and $R$ are useful here to construct stellar model for a given charge density. We also study region of strong energy and weak energy conditions in the interior of compact objects. Finally a brief discussion is given in sec. 4.

\section{Einstein- Maxwell Field Equations}
The Einstein field equation is
\begin{equation}
R_{ij} - \frac{1}{2}g_{ij} \, R = 8\pi G \;T_{ij}
\end{equation}
where $R_{ij}$, $g_{ij}$, $ R$ and $T_{ij}$ are Ricci tensor, metric tensor, Ricci scalar and energy momentum tensor respectively. We consider a  spherically symmetric static star represented by the line element,
\begin{equation}
ds^{2} = -e^{2\nu (r)}dt^{2}+e^{2\mu (r)}dr^{2}+r^{2}(d\theta
^{2}+\sin ^{2}\theta d\phi ^{2})
\end{equation}
where $\nu(r)$ and $\mu(r)$ are the two unknown metric functions. We also consider electrically charged star with most general form of energy-momentum tensor given by,
\begin{equation}
T_{ij} = \mbox{diag}~(-(\rho+E^2),~ p-E^2,~ p+E^2,~p+E^2),
\end{equation}
where $\rho$ and $p$ represents the energy density and pressure respectively and $E^2$ represents intensity of the electric field.

Using Einstein field equation, we obtain the following field equations:
\begin{equation}
\rho+E^2 = \frac{\left(1-e^{-2\mu}\right)}{r^2}+\frac{2\mu'e^{-2\mu}}{r},
\end{equation}
\begin{equation}
p-E^2 = \frac{2\nu'e^{-2\mu}}{r}-\frac{\left(1-e^{-2\mu}\right)}{r^2},
\end{equation}
\begin{equation}
p+E^2=e^{-2\mu} \left[\nu ^{\prime \prime }+{\nu ^{\prime }}^{2}-\nu ^{\prime
}\mu ^{\prime }+\frac{\nu ^{\prime }}{r}-\frac{\mu ^{\prime
}}{r}\right],
\end{equation}
where $()'$ represents derivative w.r.t. $ r$. From eqs.(5) and (6) we obtain
\begin{equation}
\nu ^{\prime \prime }+{\nu ^{\prime }}^{2}-\nu ^{\prime
}\mu ^{\prime }-\frac{\nu ^{\prime }}{r}-\frac{\mu ^{\prime
}}{r}-\frac{\left(1-e^{-2\mu}\right)}{r^2}-2E^2 e^{2\mu}=0
\end{equation}
To solve the above differential eq. (7), we use the ansatz\cite{Thomas},
\begin{equation}
e^{2\mu}=\frac{1+a r^2/R^2}{1+r^2/R^2}, 
\end{equation}
where $a$ is the spheroidicity parameter and $R$ is a geometrical parameter related with the configuration of a star model. Now using the ansatz given by eq.(8) in eq.(7), we obtain a second order differential equation in $x$, given by
\begin{equation}
(1-a+a x^2)\Psi_{xx} - a x \Psi_{x} + a(a-1)\Psi - \frac{2E^2 R^2 (1-a+ax^2)^2}{(x^2-1)}\Psi = 0 
\end{equation}
where $\Psi = e^{\nu(r)}$, with $x^{2} = 1+\frac{r^2}{R^2}$ and $\psi_{x}$ represents derivative of $\psi$ w.r.t. $x$.
\\
Now for simplicity we choose $\beta^2 = \frac{E^2 R^2(1-a+a x^2)^2}{(x^2-1)}$, such that it is related to the electric field intensity as
\begin{equation}
E^2=\frac{\beta^2(x^2-1)}{R^2(1-a+ax^2)^2}=\frac{\beta^2 r^2}{R^4 (1+\frac{ar^2}{R^2})^2}.
\end{equation}
We choose the electric field intensity as in eq. (10) so that the regularity at the centre of the compact object is ensured \cite{Tikekar}. The choice of $E^2$ generates a model for a charged star which is physically realistic in pseudo spheroidal geometry. 
Using eq. (10) and the transformation $z = \sqrt{\frac{a}{a -1}} x $, eq. (9) can be written as
\begin{equation}
(1-z^2)\Psi_{zz}+z\Psi_{z} + [(1-a)+2\beta^2/a]\Psi  = 0 
\end{equation}
Differentiating the above equation once again with respect to $z$, we get
\begin{equation}
(1-z^2)\Psi_{zzz}-z\Psi_{zz} + \Omega^2\Psi_{z}  = 0 
\end{equation}
where $\Omega^2=(2-a+2\beta^2/a)$ is a constant. Thus, for a given $\beta$, positivity of $\Omega$ puts an upper bound on the spheroidicity parameter: $a<(1+\sqrt{1+2\beta^2})$. Again in pseudo-spheroidal space-time positivity of central density puts another constraint on $a$, leading to $a>1$. Eq.(12) is a third order differential equation in $z$, we obtain the following general solutions for two different cases \cite{Tikekar}:

Case(i). $1<a<1+\sqrt{1+2\beta^2}$.
In this case  $\Omega \; (=\sqrt{2-a+2\beta^2/a})$ is positive for any $\beta$ and corresponding solution is given by
\begin{equation}
\Psi= A[\Omega\sqrt{z^2-1}\;cosh(\Omega \eta)-z\;sinh(\Omega \eta)] + B [\Omega\sqrt{z^2-1}\;sinh(\Omega \eta) - z\;cosh(\Omega \eta)]
\end{equation}

Case(ii). $a > 1+\sqrt{1+2\beta^2}$.
In this case $\Omega \; (=\sqrt{a-2-2\beta^2/a})$ is positive for all value of $\beta$, and the corresponding solution is 

\begin{equation}
\Psi= A[\Omega\sqrt{z^2-1}\;cos(\Omega\eta)-z\;sin(\Omega\eta)] + B[\Omega\sqrt{z^2-1}\;sin(\Omega\eta) + z\;cos(\Omega\eta)],
\end{equation}
where $z=cosh(\eta)$. The solutions obtained by Tikekar and Jotania\cite{Tikekar} are recovered for $\beta=0$. The unknown constants $A$ and $B$ are to be determined from the boundary conditions.\\
The total charge contained within the sphere of radius $r$ is defined as follows:
\begin{equation}
q(r)=4\pi\int_{0}^{r}\sigma r^2e^\mu dr=r^2E^2(r)
\end{equation}
where $\sigma$ denotes the proper charge density.

\section{Physical applications}

The general relativistic solutions given by eqs. $(13)$ and $(14)$ will be used in this section to study compact objects with charge. The physical parameters namely, energy density $(\rho)$, pressure  (p) and charge density $(\sigma)$ are functions of '$r$', which are given by,
\begin{equation}
\rho = {(a-1)\over R^2 (1+\frac{ar^2}{R^2})} \bigg[ 1 + {2 \over (1+\frac{ar^2}{R^2})}-\frac{\beta^2r^2}{R^2(a-1)(1+\frac{ar^2}{R^2})}\bigg],
\end{equation}

\begin{equation}
p = - {(a-1)\over R^2 (1+\frac{ar^2}{R^2})} \bigg[1-\frac{2R^2 (1+\frac{r^2}{R^2})}{r(a-1)}\frac{\Psi_{r}}{\Psi}- \frac{\beta^2 r^2}{R^2(a-1)(1+\frac{ar^2}{R^2})}\bigg],
\end{equation}

\begin{equation}
\sigma =\frac{2\beta (1+\frac{r^2}{R^2})(3+\frac{ar^2}{R^2})}{R^2(1+\frac{ar^2}{R^2})^\frac{5}{2}}.
\end{equation}
The set of equations namely eqs.(10) and eqs. (16) - (18) are relevant for determining the physical features  of a compact star. From eq. (16) it is evident that the central density of a star is independent of $\beta$ i.e. on electric field which however depends on the geometrical parameter $R$ and the spheroidicity parameter\cite{Thomas} ($a$). Thus we note that the central density of a compact star  which is given by,
\begin{equation}
\rho_{0}=\frac{3(a-1)}{R^2}
\end{equation}
is a constant for a given values of  $a$ and $R$. One interesting aspect is that it does not depend on the charge on  the compact object. Non negativity of central density ensure that $a>1$ and also in a pseudo-spheroidal geometry we need $z>0$. One recovers \cite{Tikekar}  energy density $(\rho)$ and pressure $(p)$ that for an uncharged star for $\beta=0$. In the interior of the charged sphere, energy density ($\rho$), pressure ($p$) and charge ($\sigma$) are well-behaved, bounded, finite and regular at the centre. The boundary of the star is determined from $p(b)=0$. The total mass of a charged star contained within the radius $b$ is given by,
\begin{equation}
M(b) = \frac{(a-1)b^3}{2R^2(1+a\frac{b^2}{R^2})}-\frac{\beta^2}{4a^2}\bigg[\frac{b(3+2a\frac{b^2}{R^2})}{1+a\frac{b^2}{R^2}}-\frac{3R \; tan^{-1} \left(\frac{b}{R}\sqrt{ a} \right)}{\sqrt{a}}\bigg].
\end{equation}
The compactness factor $u$ (defined as the ratio of mass to radius) is given by
\begin{equation} 
u = \frac{M(b)}{b} = \frac{(a-1)y^2}{2(1+a y^2)} - \frac{\beta^2}{4a^2}\bigg[\frac{(3+2ay^2)}{1+ay^2}-\frac{3tan^{-1}(y\sqrt a)}{y\sqrt a}\bigg]
\end{equation}
where $y = \frac{b}{R}$. In the case of a compact charged star, the following conditions are imposed:
\begin{itemize}
\item At the boundary of the star $(r=b)$, the interior solution is matched with the exterior Reissner-Nordstrom metric given by,

\begin{equation}
ds^2=-\left(1-\frac{2M}{r}+\frac{q^2}{r^2}\right)dt^2+\left(1-\frac{2M}{r}+\frac{q^2}{r^2}\right)^{-1}dr^2+r^2(d\theta^2+Sin^2\theta d\phi^2).
\end{equation}
Thus the metric potential must be matched at the boundary as follows:
\begin{equation}
e^{2\nu(r=b)} = e^{-2\mu(r=b)} = \left(1 - \frac{2M}{b}+\frac{q^2}{b^2}\right).
\end{equation}
where $M$ and $q$ denote the total mass and charge of the compact star respectively, as measured by an observer at infinity. 
\item At the boundary $(r=b)$ of the star the pressure $p=0$, which yields,
\begin{equation}
\frac{\Psi_{r}(b)}{\Psi(b)} =\frac{b(a-1)}{2R^2(1+\frac{b^2}{R^2})}\bigg[1-\frac{\beta^2b^2}{R^2(a-1)(1+\frac{ab^2}{R^2})}\bigg]
\end{equation}

\item Inside and on the surface of the charged sphere the pressure $p\geq0$, which leads to  
\begin{equation}
\frac{\Psi_{r}}{\Psi} \geq \frac{r(a-1)}{2R^2(1+\frac{r^2}{R^2})}\bigg[1-\frac{\beta^2r^2}{R^2(a-1)(1+\frac{ar^2}{R^2})}\bigg]
\end{equation}

\end{itemize}
From eqs. (16) and (17), we obtain squared of the speed of sound which is given by,
\begin{equation}
\frac{\partial p}{\partial \rho}=\frac{\frac{R^2}{r}(1+\frac{ar^2}{R^2})\bigg[\frac{R^2}{r}(1+\frac{ar^2}{R^2})(1+\frac{r^2}{R^2})(\frac{\Psi_{r}}{\Psi})^2+(a-1)(\frac{\Psi_{r}}{\Psi})\bigg]-\frac{\beta^2}{R^2}(3R^2+ar^2)}{\bigg[a(a-1)(5+\frac{ar^2}{R^2})+\frac{\beta^2}{R^2}(R^2-ar^2)\bigg]}.
\end{equation}
The speed of sound must be less than the speed of light inside the star to maintain the causality condition for which $(\frac{\partial p}{\partial\rho})<1$. Thus the causality condition leads to the following inequality equations.

\begin{equation}
\bigg(-\frac{\sqrt{a-1}}{2\sqrt{a(1+\frac{r^2}{R^2})}} - D\bigg) \leq \frac{R^2(1+\frac{ar^2}{R^2})\sqrt{1+\frac{r^2}{R^2}}}{r\sqrt{a(a-1)}}\frac{\Psi_{r}}{\Psi} \leq \bigg(-\frac{\sqrt{a-1}}{2\sqrt{a(1+\frac{r^2}{R^2})}} + D\bigg)
\end{equation}
where, $D=\sqrt{\frac{a-1}{4a(1+\frac{r^2}{R^2})}+5+\frac{ar^2}{R^2}+\frac{4\beta^2}{a(a-1)}}$.
Using inequalities (25) and (27) we obtain the following bounds on the spheroidicity parameter $a$:

\begin{equation}
(i) a < \frac{7-\sqrt{49-17(16\beta^2-3)}}{17}
\end{equation}

\begin{equation}
(ii) a > \frac{1}{2}\left[3+A+\sqrt{\frac{1}{3}(34+8\beta^2-\frac{P}{Q}-Q)+ \frac{24+8\beta^2}{A}}\right]
\end{equation} 
where $A=\sqrt{\frac{1}{3}(17+4\beta^2+\frac{P}{Q}+Q)}$, $P=25+16\beta^2+16\beta^4$ and
$Q=(125+120\beta^2+312\beta^4+64\beta^6+24\sqrt{3}a^2\sqrt{25+28\beta^2+47\beta^4+16\beta^6})^\frac{1}{3}$.

It may be pointed out here that in the uncharged case {\it i.e. } $(\beta=0)$ one obtains realistic solutions for  (i) $a<-\frac{3}{17}$ and (ii) $a>5$ respectively\cite{PKC}. In the pseudo-spheroidal
 geometry positivity of the spheroidal parameter $a$  is ensured for $\beta\geq0$. In this case, the maximum compactness factor upto which a compact stellar model permitted is $u=0.4167$ for $\beta=0$ with $a=6$ . However, we note that for $\beta=0.65$, the spheroidicity parameter satisfy a lower bound ($a>5.16208$) for the same compactness factor $u=0.4167$.

\subsection{Numerical results}

The energy density and pressure of a compact star  are determined by parameters $a$, $R$ and $\beta$. It is not simple to obtain a known form of pressure $(p)$ in terms of energy density $(\rho)$ as these are highly non-linear functions of those parameters. So we adopt numerical technique in the next sections to study the physical properties of compact objects. We plot the radial variation of different physical parameters such as $p$, $(\rho-p)$, $(\rho-3p)$ and ($\frac{\partial p}{\partial\rho}$) for different $a$, $\beta$ and $R$. We tabulate radial variation of energy density with various $a$ and $\beta$. To construct a stellar model, first we determine $R$ using eqs. (8) and (23) in the uncharged limit for a given configuration of compact object namely, the spheroidicity parameter $(a)$, Mass $ (M)$ and size $(b)$. To study the dependence of charge on the different physical parameters, we determine the two unknown constants $A$ and $B$ of eqs.(13) and (14). It is possible to determine $A$ and $B$ from eqs.(23) and (24) for a particular choice of mass $(M)$, radius $(b)$, spheroidicity parameter $(a)$ and charge parameter $(\beta)$. Once $A$, $B$ and $R$ are known, the radial variation of the physical parameters namely, energy  density ($\rho$), pressure  ($p$), $(\rho-p)$, $(\rho-3p)$ and $(\frac{\partial p}{\partial\rho})$ are determined using eqs.(16) and (17) for different $\beta$.  The compactness factor $u=\frac{M}{b}$ is determined for a given spheroidicity parameter $a$ using eq.(21). In the next section, we consider three different cases to obtain stellar models of known compact stars, where the masses are taken from observations\cite{Latt}.

{\bf Case I:} Using the data for X-ray pulsar, namely Her X-1\cite{RS01} characterised by Mass $M = 0.88M_{\odot}$, where $M_{\odot}$ = the solar mass, size of the star $b = 7.7$~km we obtain stellar model. It has compactness factor $u=M/b=0.1686$. The geometrical parameter, namely, the spheroidicty parameter in this case is $a=6$.

\begin{figure}
\begin{center}
\includegraphics{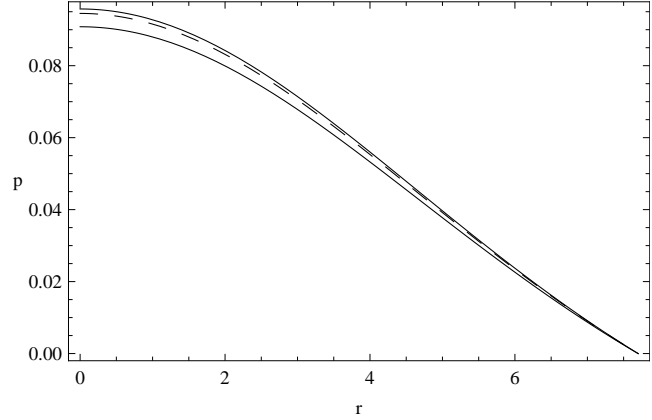}
\caption{Radial variation of pressure($p$) in $GeV/fm^3$ inside the star HER X-1. Curves from top to bottom are for $\beta=0, 0.3, 0.6$ respectively.}
\end{center}
\end{figure}
The above values are then used to determine the geometrical parameter $R$ from eqs. (8) and (23) at $r=b$. For a star without charge {\it i.e.}  $\beta=0$, we obtain  $R=22.882$~km.  However, for a charged star with  $\beta=0.3$ and $\beta= 0.6$, we get a viable compact star model for $R= 22.905$ km. and $R= 22.974$ km.  respectively. The radial variation of pressure ($p$) for these models are shown in fig. (1) for  $\beta=0, 0.3, 0.6$.

\begin{figure}[h!]
\begin{center}
\includegraphics{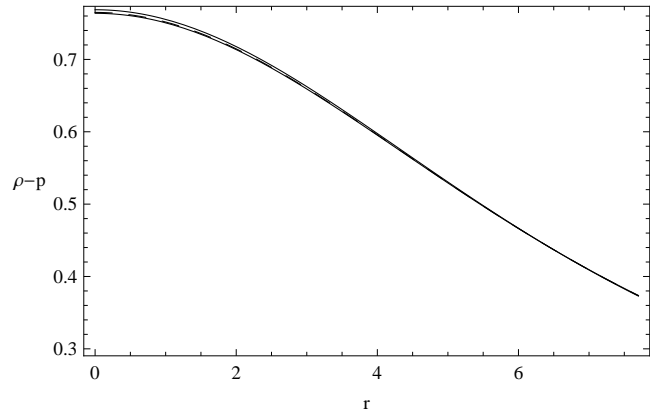}
\caption{Variation of $(\rho-p)$ in $GeV/fm^3$ for  HER X-1 with radial distance $r$ in km. Curves from top to bottom are for $\beta=0.6, 0.3, 0$ respectively.}
\end{center}
\end{figure}

It is   evident from fig.(1) that the radial pressure ($p$) decreases with an increase in charge which is prominent at the centre. In table-1 radial variation of energy density ($\rho$) in the interior and on the surface of the star are tabulated for two sets of parameters. It is evident that the energy density ($\rho$) decreases with an increase in charge inside  the compact star except at the centre for a given spheroidicity parameter. It is also evident that for a large spheroidicity parameter the radial variation of energy density is same in  all the cases.   In fig.(2) and fig.(3) radial variation of $(\rho-p)$ and $(\rho-3p)$ are plotted to determine the regions where  weak energy condition $(\rho-p)$ (henceforth, WEC) and strong energy condition $(\rho-3p)$ (henceforth SEC) separately are obeyed. Although radial variation of $(\rho-p)$ overlaps, a significant variation in $(\rho-3p)$ is observed when the charge is increased. The variation of $(\frac{\partial p}{\partial\rho})$ at the centre and at the surface for different charge parameter $\beta$ are plotted in  fig.(4). It is evident that the causality condition ($\frac{\partial p}{\partial\rho}<1$) holds good inside the star even in the presence electric field.  

\begin{figure}
\begin{center}
\includegraphics{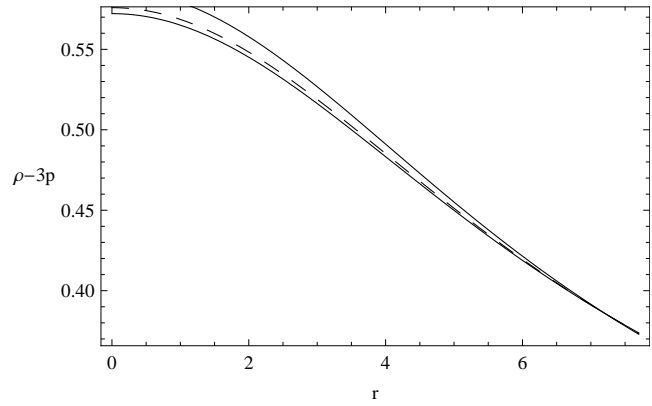}
\caption{Variation of $(\rho-3p)$ in $GeV/fm^3$ for HER X-1 with the radial distance $r$ in Km. curves  from top to bottom are for $\beta=0.6, 0.3, 0$ respectively.}
\end{center}
\end{figure}

\begin{figure}[h!]
\begin{center}
\includegraphics{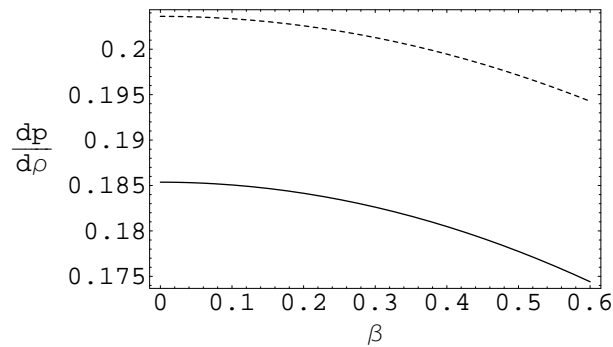}
\caption{Variation of $(\frac{\partial p}{\partial\rho})$ with electric charge  (i.e. $\beta$) at the center (Solid line) and at the surface (Dotted line) of the star HER X-1.}
\end{center}
\end{figure}

\begin{table}
\begin{center}
\begin{tabular}{|c|c|c|c|c|c|c|}  \hline
\multicolumn{1}{|c|}{} & \multicolumn{6}{|c|}{Energy density($\rho$)} \\ \cline{2-7}
\multicolumn{1}{|c|}{$r$ in km.}    & \multicolumn{3}{|c|}{$a=6, R=22.882$ km.} & \multicolumn{3}{|c|}{$a=15, R=39.657$ km.}\\ 
\cline{2-7}
           & $\beta=0$    & $\beta=0.3$  & $\beta=0.6$  & $\beta=0$  & $\beta=0.3$   & $\beta=0.6$ \\ \hline
0          & 0.85949      & 0.85949      & 0.85949      & 0.80117    & 0.80117       & 0.80117     \\ \hline
2.0        & 0.79781      & 0.79777      & 0.79767      & 0.75283    & 0.75282       & 0.75281     \\ \hline
4.0        & 0.65129      & 0.65118      & 0.65084      & 0.63374    & 0.63373       & 0.63369     \\ \hline
b          & 0.37374      & 0.37353      & 0.37291      & 0.38853    & 0.38850       & 0.38842     \\ \hline
\end{tabular}
\end{center}
\label{tab1}{Table-1 : Energy density $(\rho)$ in $GeV/fm^3$ in the interior and on the surface of the compact star X-ray pulser HER X-1 having  Mass $M=0.88M_{\odot}$ and Radius $b=7.7$~km. with compactness $u=0.1686$.}
\end{table}

In the next section, we consider known compact objects with two different configurations. First let us consider SAX J with two possible models of compactness\cite{Tikekar,RS02}.
 
{\bf Case IIa:} we consider here a millisecond pulsar namely, SAX J 1808.4-3658 which is characterised by mass $M=1.435M_{\odot}$, and size of the star $b = 7.07$~km. We note that the above star configuration is permitted for compactness $u=M/b=0.2994$ with spheroidicity parameter $a=6$. Using eq.(8) along with the boundary condition at $r=b$, we determine geometrical parameter $R=10.839$~km. 

\begin{figure}[h!]
\begin{center}
\includegraphics{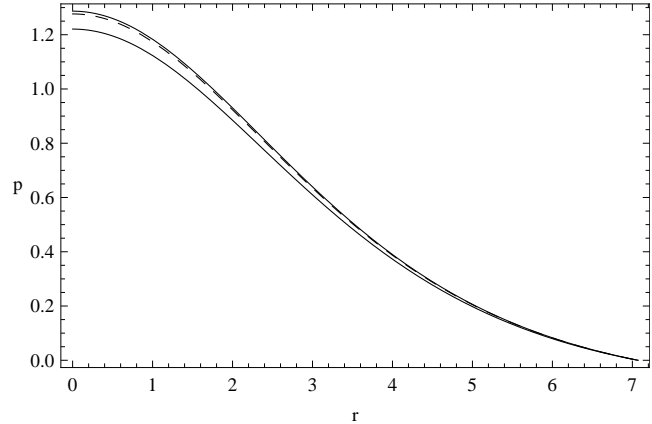}
\caption{Variation of pressure $p$ in $GeV/fm^3$ with the radial distance $r$ in km inside the star SAX J. Curves from top to bottom are for $\beta=0, 0.2351, 0.6$ respectively.}
\end{center}
\end{figure}

\begin{figure}
\begin{center}
\includegraphics{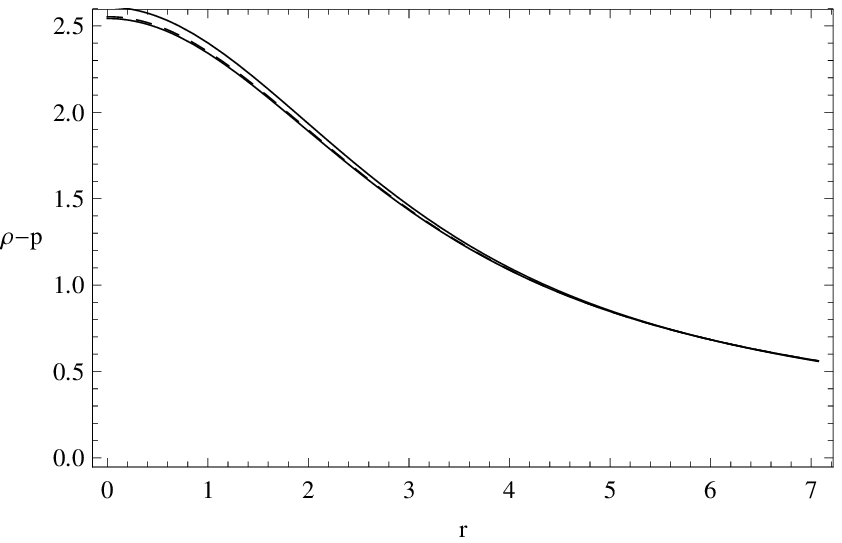}
\caption{Variation of ($\rho-p$) in $GeV/fm^3$ inside the star SAX J with the radial distance $r$ in =km. Curves from top to bottom are for $\beta=0.6, 0.2351, 0$ respectively.}
\end{center}
\end{figure}

\begin{figure}[h!]
\begin{center}
\includegraphics{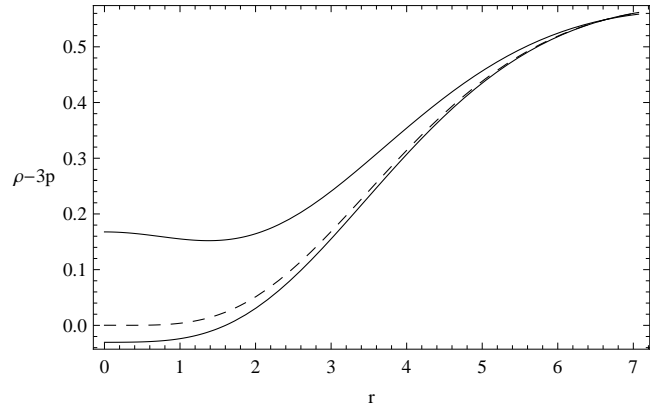}
\caption{Variation of ($\rho-3p$) in $GeV/fm^3$ inside the star SAX J with the radial distance $r$ in km.  Curves from top to bottom are for $\beta=0.6, 0.2351, 0$ respectively.}
\end{center}
\end{figure}

Using the parameters we determine its various physical properties qualitatively. The radial variation of pressure ($p$) is plotted in  fig.(5) for given electric charges ({\it i.e.,} $\beta$). It is found that the effect of charge is to reduce the pressure inside the star. It is evident from fig.(6) that WEC $(\rho-p) > 0 $  is also obeyed. The energy density ($\rho$) is tabulated with radial distance in table-2 for two  sets of parameters $a=6$ and $a=15$.   We plot the radial variation of  $(\rho-3p)$ in fig.(7) both for charged and uncharged compact stars. In the case of uncharged star it is found that the SEC  is violated near the centre but gradually away from the centre  $(\rho-3p)$ attains a positive value. In the case of charged star with $\beta \geq 0.2351$ SEC condition is maintained throughout the interior of compact object.
Thus for a compact star with or without charge, we note an interesting case where a compact star may be assumed to be made up of a core and an envelope region. In the case of uncharged star SAX J1 with $a=6$, a core region upto a sphere  of radius $r= 1.602~$ km. might exit. In the case of a charged star with $\beta = 0.15$, one obtains a core of radial distance $r= 1.412~km.$ from the centre where normal perfect fluid may not exist. It is also observed that the core region  decreases due to the presence of charge on the compact object.  For $\beta\geq0.2351$,   such a core region does not exist as  SEC condition is obeyed inside at all region of the star.      This is an interesting case, where one obtains an unusual behaviour of matter inside the compact star for low value of charge density which remains to be explored elsewhere. In fig.(8) variation of ($\frac{\partial p}{\partial\rho}$) with $\beta$ at the centre and at the surface of star SAX J-1 is shown. The causality condition  $(\frac{\partial p}{\partial\rho}<1)$ is maintained in  the interior of the star.

\begin{figure}
\begin{center}
\includegraphics{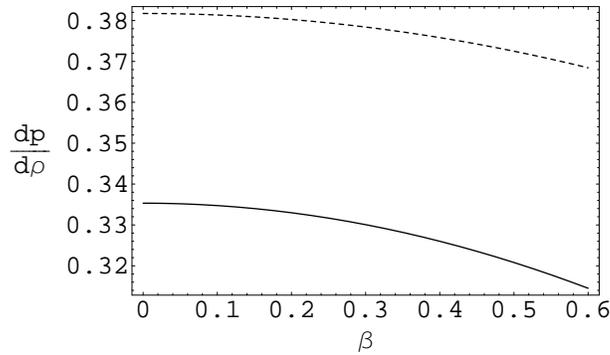}
\caption{Variation of $(\frac{\partial p}{\partial\rho})$ with charge $\beta$ at the center (Solid line) and at the surface (Dotted line) for  SAX J-1.}
\end{center}
\end{figure}  

\begin{table}
\begin{center}
\begin{tabular}{|c|c|c|c|c|c|c|}  \hline
\multicolumn{1}{|c|}{} & \multicolumn{6}{|c|}{Energy density($\rho$)} \\ \cline{2-7}
\multicolumn{1}{|c|}{$r$ in km.} &\multicolumn{3}{|c|}{$a=6$, R=10.8394~km.}  & \multicolumn{3}{|c|}{$a =15$, R=20.4683~km.} \\ \cline{2-7}
           & $\beta=0$ & $\beta=0.3$  & $\beta=0.6$  & $\beta=0$ & $\beta=0.3$  & $\beta=0.6$  \\ \hline
0          & 3.83003   & 3.83003      & 3.83003      & 3.00751   & 3.00751      & 3.00751      \\ \hline
2.0        & 2.82074   & 2.82020      & 2.81858      & 2.41104   & 2.41099      & 2.41085      \\ \hline
4.0        & 1.47593   & 1.47498      & 1.47214      & 1.44784   & 1.44775      & 1.44745      \\ \hline
b          & 0.56168   & 0.56090      & 0.55858      & 0.61701   & 0.61691      & 0.61661      \\ \hline
\end{tabular}
\end{center}
\label{tab2}{Table-2: Energy density $(\rho)$ in $GeV/fm^3$ in the interior and on the surface of the compact star SAX-J having parameters Mass $M=1.435M_{\odot}$ and Radius $b=7.07$~km. with compactness $u=0.2994$.}
\end{table}

{\bf Case IIb:} We now consider another model for the same star SAX J1 which is characterised by mass $M = 1.323M_{\odot}$ and radius $b = 6.55$~km. The above configuration can be accommodated with compactness factor $u=M/b=0.2979$ in our model with spheroidicity parameter $a = 6$. In this case we consider a star with smaller compactness factor than  that taken in Case IIa.

\begin{figure}
\begin{center}
\includegraphics{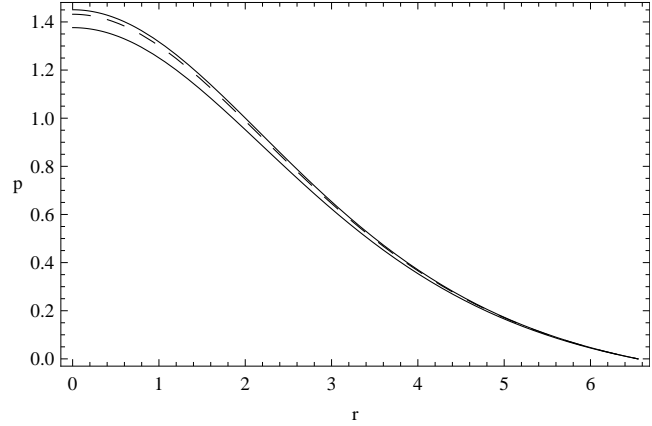}
\caption{Variation of pressure $p$ in $GeV/fm^3$ inside the star SAX J with the radial distance $r$ in km. for $M = 1.3235M_{\odot}$, radius $b = 6.55$~km. with  $a = 6$ ($R=10.1288$~km). Curves from top to bottom are for $\beta=0, 0.3, 0.6$ respectively.}
\end{center}
\end{figure}

\begin{figure}[h!]
\begin{center}
\includegraphics{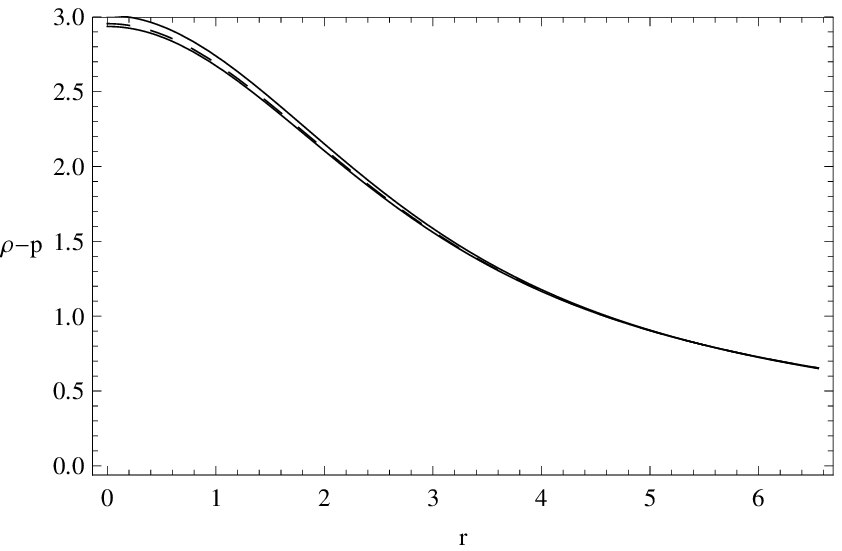}
\caption{Variation of ($\rho-p$) in $GeV/fm^3$ inside the star SAX J with the radial distance $r$ in Km. Curves from top to bottom are for $\beta=0.6, 0.3, 0$ respectively.}
\end{center}
\end{figure}

\begin{figure}
\begin{center}
\includegraphics{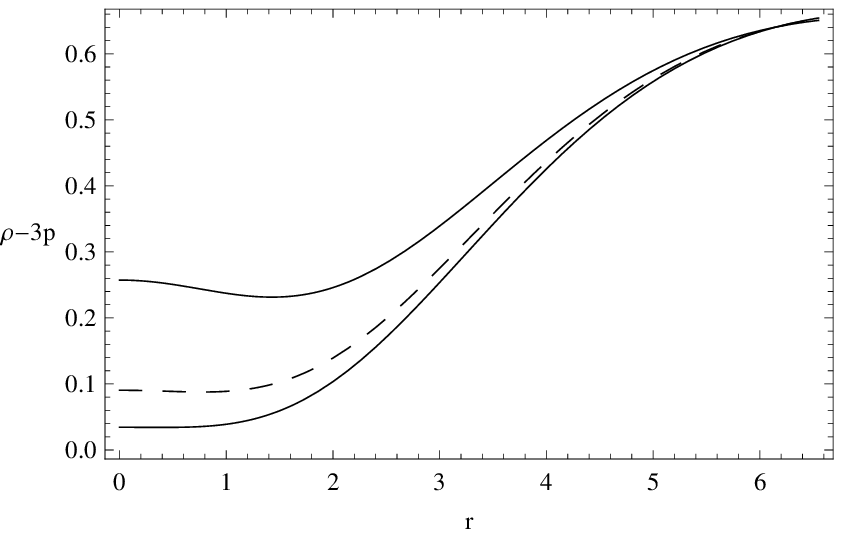}
\caption{Variation of ($\rho-3p$) in $GeV/fm^3$ inside the star SAX J with the radial distance $r$ in km. Curves from top to bottom are for $\beta=0.6, 0.3, 0$ respectively.}
\end{center}
\end{figure}
The geometrical parameter $R$ is determined from the eqs.(8) and (23) using the boundary condition at $r = b$, which is $R=10.129$~km.  The radial variation of pressure ($p$) for different charge parameters are shown in fig.(9).  The radial variation of energy density $(\rho)$ is tabulated in table-3. From figs.(10) and (11), it is evident that both the weak and strong energy conditions are obeyed inside  the star with or without  electric charge. Fig.(12) shows the variation of ($\frac{\partial p}{\partial\rho}$) with $\beta$ at the centre and in the surface of star SAX J-1. The causality condition $(\frac{\partial p}{\partial\rho}<1)$ is obeyed inside the star for the chosen set of parameters. In table-4, we tabulated $(\rho-p)$ and $(\rho-3p)$ at the centre of the star for spheroidicity parameter $a=6$ with different compactness factor ($u$). It is found that at higher compactification factor there exists  regions inside the star where not only SEC is violated but also WEC. We note from columns 2 and 3 of table-4 that in the uncharged case strong energy condition $(\rho-3p)\geq0$ is violated when compactness factor of the star is  $u>0.29866$ whereas weak energy condition $(\rho-p)\geq0$ is violated when $u>0.35615$. 

\begin{figure}
\begin{center}
\includegraphics{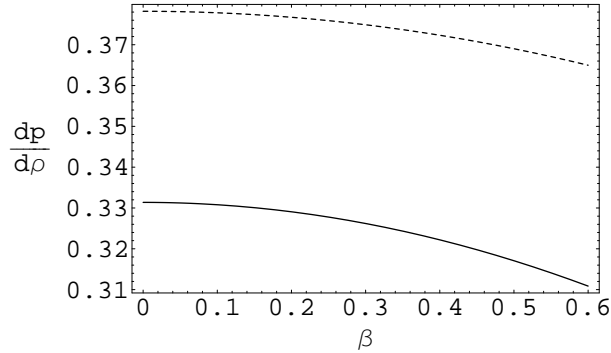}
\caption{Variation of ($\frac{\partial p}{\partial\rho}$) with charge parameter $\beta$ at the center (Solid line) and at the surface (Dotted line) of the star SAXJ-1.}
\end{center}
\end{figure}

\begin{figure}[h!]
\begin{center}
\includegraphics{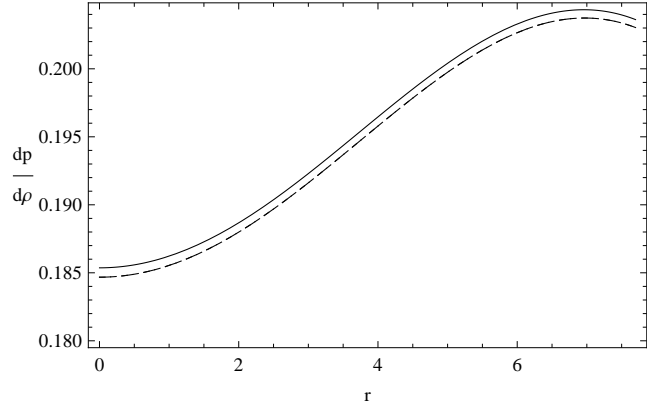}
\caption{Variation of ($\frac{\partial p}{\partial\rho}$) with  $r$ for the star HER X1.}
\end{center}
\end{figure}

\begin{figure}[h!]
\begin{center}
\includegraphics{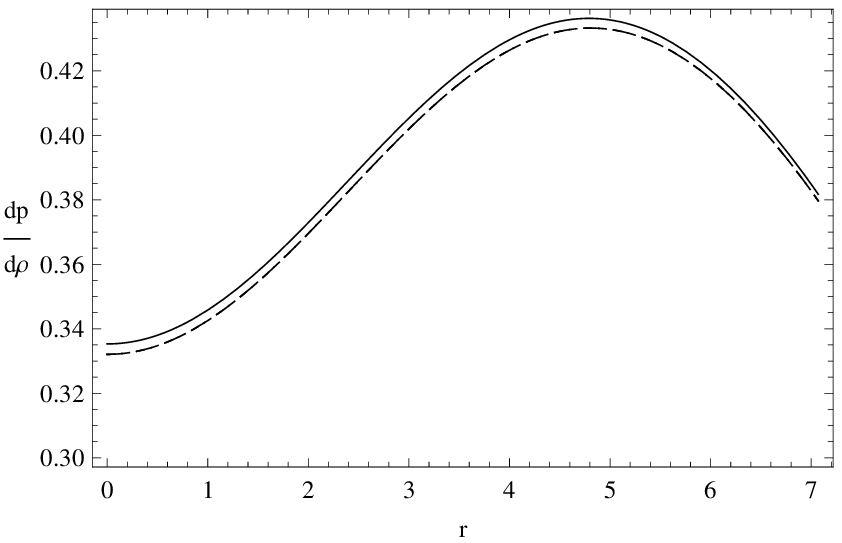}
\caption{Variation of ($\frac{\partial p}{\partial\rho}$) with  $r$ for the star SAXJ-1 (Case-IIa).}
\end{center}
\end{figure}

\begin{figure}[h!]
\begin{center}
\includegraphics{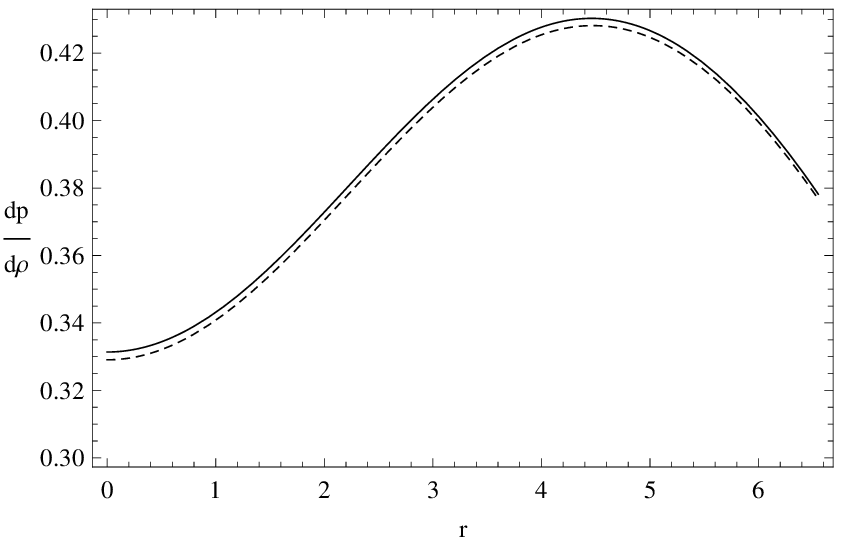}
\caption{Variation of ($\frac{\partial p}{\partial\rho}$) with  $r$ for the star SAXJ-1(Case-IIb).}
\end{center}
\end{figure}

\begin{figure}[h!]
\begin{center}
\includegraphics{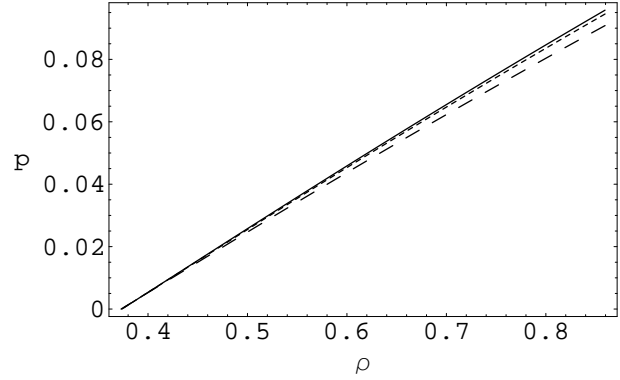}
\caption{Variation of $p$ with $\rho$ for HER X1 (Case-I).}
\end{center}
\end{figure}

\begin{figure}
\begin{center}
\includegraphics{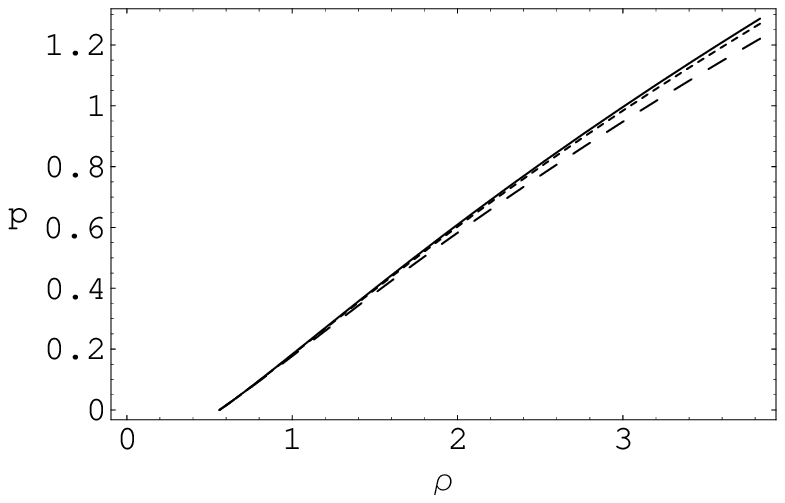}
\caption{Variation of $p$ with $\rho$ for  SAXJ-1 (Case-IIa).}
\end{center}
\end{figure}

\begin{figure}
\begin{center}
\includegraphics{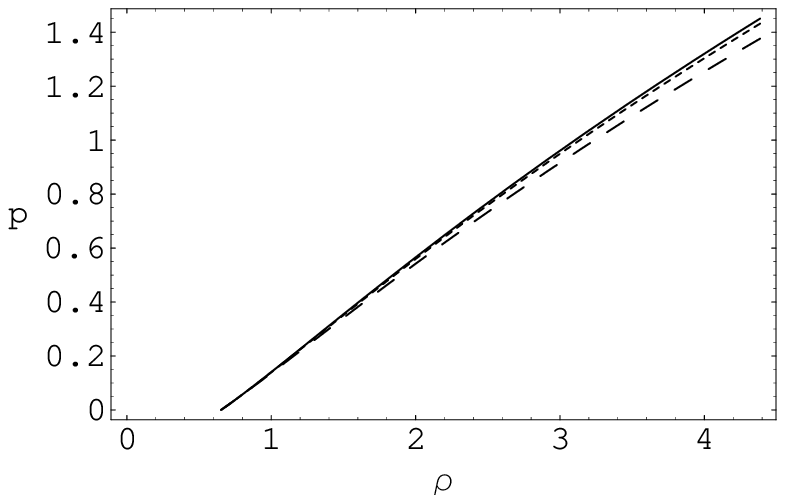}
\caption{Variation of $p$ with $\rho$ for SAXJ-1 (Case-IIb).}
\end{center}
\end{figure}

\begin{table}
\begin{center}
\begin{tabular}{|c|c|c|c|c|c|c|}  \hline
\multicolumn{1}{|c|}{} & \multicolumn{6}{|c|}{Energy density($\rho$)} \\ \cline{2-7}
\multicolumn{1}{|c|}{$r$ in Km.} & \multicolumn{3}{|c|}{$a=6, R=10.129$~km.} & \multicolumn{3}{|c|}{$a=15, R=19.092$~km.}   \\ \cline{2-7}
           & $\beta=0$ & $\beta=0.3$  & $\beta=0.6$  & $\beta=0$ & $\beta=0.3$  & $\beta=0.6$  \\ \hline
0          & 4.38628   & 4.38628      & 4.38628      & 3.45693   & 3.45693      & 3.45693      \\ \hline
2.0        & 3.10543   & 3.10476      & 3.10274      & 2.68859   & 2.68853      & 2.68835      \\ \hline
4.0        & 1.53570   & 1.53461      & 1.53132      & 1.53270   & 1.53258      & 1.53222      \\ \hline
b          & 0.65413   & 0.65324      & 0.65056      & 0.71797   & 0.71786      & 0.71751      \\ \hline
\end{tabular}
\end{center}
\label{tab3}{Table-3: Energy density $(\rho)$ in $GeV/fm^3$ in the interior and on the surface of the compact star SAX-J having mass $M=1.323M_{\odot}$ and radius $b=6.55$~km. with compactness $u=0.2979$.}
\end{table}

\begin{table}
\begin{center}
\begin{tabular}{|c|c|c|c|c|c|c|}  \hline
\multicolumn{1}{|c|}{u} & \multicolumn{2}{|c|}{$\beta=0$} & \multicolumn{2}{|c|}{$\beta=0.3$} & \multicolumn{2}{|c|}{$\beta=0.6$}   \\ \cline{2-7}
           & $\rho-p$  & $\rho-3p$    & $\rho-p$     & $\rho-3p$ & $\rho-p$     & $\rho-3p$    \\ \hline
0.29800    & 10.03540  & 0.10608      & 10.09930     & 0.29794   & 10.28970     & 0.86902      \\ \hline
0.29865    & 10.00030  & 0.00085      & 10.06480     & 0.19427   & 10.25660     & 0.76994      \\ \hline
0.29866    & 9.99974   & -0.00078     & 10.06420     & 0.19266   & 10.25610     & 0.76841      \\ \hline
0.29990    & 9.93147   & -0.20558     & 9.99697      & -0.00908  & 10.19190     & 0.57567      \\ \hline
0.30340    & 9.72879   & -0.81364     & 9.79737      & -0.60790  & 10.00130     & 0.00394      \\ \hline
0.30345    & 9.72578   & -0.82267     & 9.79440      & -0.61680  & 9.99848      & -0.00455     \\ \hline
0.35614    & 0.00433   & -29.9870     & 0.32137      & -29.0359  & 1.22816      & -26.3155     \\ \hline
0.35615    & -0.00202  & -30.0060     & 0.31525      & -29.0542  & 1.22268      & -26.3220     \\ \hline
0.35650    & -0.22879  & -30.6864     & 0.09679      & -29.7096  & 1.02708      & -26.9188     \\ \hline
0.35670    & -0.36200  & -31.0860     & -0.0315      & -30.0944  & 0.91237      & -27.2629     \\ \hline
0.35817    & -1.43218  & -34.2965     & -1.0605      & -33.1815  & -0.00418     & -30.0125     \\ \hline
\end{tabular}
\end{center}
\label{tab4}{Table-4: Values of ($\rho-p$) and ($\rho-3p$) at the centre of the compact object for  different compactness factor and charge parameter $\beta$ in the unit of $(\frac{30}{R^{2}}) GeV/fm^{3}$ with $a=6$.}
\end{table}

\begin{table}
\begin{center}
\begin{tabular}{|c|c|c|c|c|}  \hline
\multicolumn{5}{|c|}{$\bar{b}=\frac{b}{R}$} \\ \hline
$ a $   & $\beta=0$   & $\beta=0.2$   & $\beta=0.4$  &  $\beta=0.6$      \\ \hline
6           & 2.1517      & 2.1663        & 2.2120       &  2.2935           \\ \hline
7           & 1.5047      & 1.5090        & 1.5222       &  1.5449           \\ \hline
8           & 1.2194      & 1.2214        & 1.2276       &  1.2380           \\ \hline
9           & 1.0503      & 1.0515        & 1.0550       &  1.0609           \\ \hline
10          & 0.9356      & 0.9363        & 0.9386       &  0.9423           \\ \hline
11          & 0.8513      & 0.8518        & 0.8534       &  0.8560           \\ \hline
15          & 0.6542      & 0.6543        & 0.6548       &  0.6557           \\ \hline
50          & 0.3042      & 0.3042        & 0.3043       &  0.3043           \\ \hline
100         & 0.2089      & 0.2088        & 0.2088       &  0.2088           \\ \hline
\end{tabular}
\end{center}
\label{tab5}{Table-5: The variation of $\bar{b}=(\frac{b}{R})$ with $a$ for different charge configurations $\beta$.}
\end{table}

\begin{table}
\begin{center}
\begin{tabular}{|c|c|c|c|c|c|}  \hline
\multicolumn{1}{|c|}{Star} & \multicolumn{1}{|c|}{Mass(m)} & \multicolumn{1}{|c|}{$a$} & \multicolumn{3}{|c|}{Star Size(b) in km.}\\ \cline{4-6} \multicolumn{1}{|c|}{} & \multicolumn{1}{|c|}{}& \multicolumn{1}{|c|}{}  & $\beta=0$ & $\beta=0.2$ & $\beta=0.4$                        \\ \hline
HER X-1    & 0.88$M_{\odot}$    & 6            & 7.7000       & 7.7134   & 7.7543         \\ \cline{3-6}
           &                    & 15           & 7.7000       & 7.7030   & 7.7121         \\ \hline
SAX J      & 1.435$M_{\odot}$   & 6            & 7.0700       & 7.0748   & 7.0892         \\ \cline{3-6}
SS1        &                    & 15           & 7.0700       & 7.0710   & 7.0740         \\ \hline
SAX J      & 1.323$M_{\odot}$   & 6            & 6.5500       & 6.5544   & 6.5678         \\ \cline{3-6}
SS2        &                    & 15           & 6.5500       & 6.5510   & 6.5537         \\ \hline
\end{tabular}
\end{center}
\label{tab6}{Table-6 : Size of the star(b)  with spheroidicity parameters $a=6$ and $a=15$.}
\end{table}

But in the case of a charged star  with $\beta=0.3$, strong energy condition (SEC) is violated when $u>0.29990$ whereas weak energy condition (WEC) is violated when $u>0.35670$, which are shown in columns 4 and 5 of table-4. Thus if one prescribes 3-pseudo spheroidal geometry for the 3-space of the interior space-time of compact objects, then it comes out from the analysis that there exist considerable regions with different parameters where both the WEC and SEC are violated. For a given $a$, as the charge (electric field intensity) is increased, the region of violation of WEC and SEC are found to occur at higher compactness factor for the same spheroidicity parameter ($a$).  

The upper limit on the boundary of a star is obtained  from eq.(29) which is given by
\begin{equation}
\frac{b}{R}=\sqrt{\frac{(12a^2-12a+2a\beta^2+6\beta^2)+\epsilon}{a^4-6a^3+5a^2-2a^2\beta^2+2a\beta^2+\beta^4}}
\end{equation}
where $\epsilon=(-12a^2+12a-2a\beta^2-6\beta^2)^2+(a^4-6a^3+5a^2-2a^2\beta^2+2a\beta^2+\beta^4)(17a^2-14a+16\beta^2-3)$.
In the uncharged case we obtain a limiting value same as that  obtained in the Ref. \cite{PKC}. However we note that in the presence of charge the limiting values of the reduced radius is increased for same spheroidicity parameter which are tabulated in table-5.

Finally we determine the radius of a star for a given set of parameters in the presence and in the absence of charge. The size of a compact star is tabulated in table-6. It is evident that in the presence of charged star the size of a star ($b$) is increased.

\section{Discussion}

In this paper we obtain a class of relativistic solution for static compact  star with charge in the framework of pseudo-spheroidal space-time geometry. The interior solutions of Einstein-Maxwell equation for a charged compact object particularly with electric field is matched with Reissner-Nordstrom metric at the boundary. The interior geometry of a compact star here contains five parameters namely, mass $(M)$, radius $(b)$, spheroidicity parameter ($a$), charge $(\beta)$ and geometrical parameter $(R)$. The solutions given by  eqs. (13) and (14) obtained from the Einstein's field equation contains two unknown constants namely, $A$ and $B$. These constants are determined from the two boundary conditions: (i) pressure at the boundary of the star vanishes and (ii) by matching the first fundamentals of metrics at the boundary with Reissner-Nordstrom metric. The radial variation of density, pressure and charge configuration are then determined. We note the following: 

(i) Fig-1 shows the variation of inner pressure for a given star namely, HER-X1. Since the equation of state of the matter content in the compact object is not known, we use the solutions of the Einstein field equations corresponding to a particular geometry  namely, Vaidya-Tikekar metric to estimate the energy density and pressure. The observed values of mass and radius  of HER X1 are taken to study the physical properties of the star with  pseudo-spheroidal geometry. It is found that for a given spheroidicity parameter, the pressure inside the star is less compared to that of an uncharged star. With an increase in charge $(q)$, the pressure inside the star is found to  decrease. 

(ii) The variation of pressure at the center is maximum which gradually decreases away from the center and finally vanishes at the boundary. However in the case of energy density for a given star with different  $R$ parameter the central density is found to decrease with charge.  At the  surface, energy density ($\rho$) depends on $R$ but independent of charge on the star. We note that both the weak and strong energy conditions are  obeyed inside the star as shown in fig. (2) and fig. (3) respectively. The causality condition is also obeyed throughout the star from the centre to the boundary (see fig.(4)). 

(iii) In the case of SAX J1 808.4-3658 we obtain stellar models with two different compactness factor. The radial variation of ($\rho-3p$) is shown in fig. (7). We found some interesting regions  near the centre from the graph where both SEC and WEC are violated. It is evident that for $\beta\geq0.2351$, both the strong and weak energy conditions are obeyed with compactification factor $u=0.2994$. However, for $\beta < 0.2351$ the strong energy condition is violated near the centre of the star. The radial variation of ($\rho-3p$)  is shown in fig.(11). It is evident that strong energy condition  ($\rho-3p\geq0$) is obeyed  inside the star with compactness $u=0.2979$ (Case-IIb). We tabulated the variation of  ($\rho-3p$) and ($\rho-p$) with compactification  factor $u$ for different charge parameter $\beta$. Thus as $u$ increases both WEC  and SEC violation region is found to increase both in the absence and in the presence of charge. From  figs. (8) and (12), it is evident  that causality condition is obeyed inside the star in the case of SAX J for both the  models considered here. In case of SAX J with compactness $u=0.2994$, the strong energy condition $(\rho-3p\geq0)$ is violated within a sphere of radius $r=1.602$~km. 

For a charged compact star (Case-IIa) with $\beta=0.15$, SEC violation region is found to exist within a sphere of radius of $r=1.412$~km. However for $\beta>0.2351$, such region does not exist. In this case we found a lower bound on $E^2$ for which SEC and WEC hold good. We note that  $q(r)=r^2E^2\geq\frac{4.00391\times10^{-6}r^4}{(1+0.051067r^2)^2}$ when $\beta=0.2351$ with $a=6$. However the radial variation of pressure and density are found same  as was obtained in the case of HER X-1. It is also evident that energy density is maximum at the center which decreases radially outward from the center. The energy density $(\rho)$ of a charged compact star decreases compared to that of an uncharged compact object\cite{VT}   except at the centre. In the case of higher  spheroidicity parameter $(a)$ the effect of charge in deciding physical properties of a star  is negligible. The variation of $\frac{\partial p}{\partial\rho}$ with radial distance ($r$) for case-I, Case-IIa and case-IIb are shown in figs.(13)-(15) respectively which show that the  causality is maintained inside the charged compact star. The variation of pressure with density is plotted in figs.(16)-(18), it is evident that EOS for a star without charge is linear. However, for a compact charged star the EOS is  non-linear. A deviation from linearity is visible due to non zero $\beta$.

\section{Acknowledgement:}

PKC would like to thank University Grants Commission, Eastern India for awarding a project. Authors are thankful to IUCAA Resource Centre for extending facilities to carryout research work at Physics Department, North Bengal University. Finally, BCP would like to thank IUCAA, Pune for awarding Visiting Associateship to initiate the work.

\pagebreak

\end{document}